\documentclass[prb,aps,twocolumn,showpacs,amsmath,amssymb,superscriptaddress,floatfix]{revtex4-2}

\usepackage{graphicx}
\usepackage{dcolumn}
\usepackage{bm}
\usepackage{amssymb,mathtools}
\usepackage{booktabs}
\usepackage{color}
\usepackage{microtype}
\usepackage[T1]{fontenc}
\usepackage{lmodern} 
\usepackage{enumitem}
\usepackage{times}

\usepackage{physics}
\usepackage{upgreek}

\allowdisplaybreaks
\usepackage{placeins}

\usepackage[colorlinks,plainpages=false,linkcolor=blue,urlcolor=blue,citecolor=blue,pdfpagemode=UseNone,pdfstartview=FitBH]{hyperref}

\usepackage{sidecap}

\renewcommand{\tablename}{Table}
\makeatletter\renewcommand{\fnum@table}[1]{\tablename~\thetable.}\makeatother

\makeatletter

\def\titlecase#1{%
\let\tc@w\@empty
\protected@edef\tmp{\noexpand\tc@a#1\relax}\expandafter\tc@uc@\tmp}

\def\tc@a{\futurelet\tmp\tc@aa}

\def\tc@aa{%
\ifcat a\noexpand\tmp\expandafter\tc@ab
\else\expandafter\tc@ac\fi}

\def\tc@ab#1{\edef\tc@w{\tc@w#1}\tc@a}

\def\tc@ac{%
\csname tc@@\tc@w\endcsname\expandafter\tc@uc\tc@w
\let\tc@w\@empty
\ifx\tmp\@sptoken\let\next\tc@sp
\else\ifx\tmp\relax\let\next\relax
\else\let\next\tc@nxt
\fi\fi\next}

\def\tc@sp#1{ \tc@a#1}
\def\tc@nxt#1{#1\tc@a}

\def\tc@uc#1{\uppercase{#1}}
\def\tc@uc@#1#2{\uppercase{#1#2}}

\let\tc@@the\@gobbletwo
\let\tc@@and\@gobbletwo

\makeatother

\newcommand{\mystrut}{\rule{0pt}{4ex}}
\newcommand{\ii}{\mathrm{i}}
\renewcommand{\Re}{\operatorname{Re}}
\renewcommand{\Im}{\operatorname{Im}}

\usepackage[normalem]{ulem}
\usepackage{xcolor}

\begin{document}

\title{Nonequilibrium Dynamics in a Quantum Spin Chain with Pump-Probe Resonant Inelastic X-ray Scattering}

\author{Menglian Shi}
\affiliation{National Laboratory of Solid State Microstructures and Department of Physics, Nanjing University, Nanjing 210093, China}

\author{Zhengzhong Du}
\affiliation{National Laboratory of Solid State Microstructures and Department of Physics, Nanjing University, Nanjing 210093, China}

\author{Yi Lu}
\email{yilu@nju.edu.cn}
\affiliation{National Laboratory of Solid State Microstructures and Department of Physics, Nanjing University, Nanjing 210093, China}
\affiliation{Collaborative Innovation Center of Advanced Microstructures, Nanjing University, Nanjing 210093, China}

\begin{abstract}
We investigate the nonequilibrium dynamics of the transverse field Ising chain (TFIC) using time-resolved resonant inelastic X-ray scattering (tr-RIXS) in a pump-probe setup within a condensed matter setting. The tr-RIXS spectra reveal distinctive features, including two-kink continua at high energies and oscillatory spectral weight at low energies, both strongly influenced by the pump dynamics. By systematically analyzing the nonequilibrium tr-RIXS cross section under various pump configurations, we demonstrate how the spectra encode detailed information about the system’s transient states, capturing instantaneous Hamiltonian parameters and the traversal of equilibrium quantum critical points during dynamic evolution. Notably, we uncover a one-to-one correspondence between oscillatory features in the low-energy spectra and dynamical quantum phase transitions, offering a novel and experimentally accessible approach for their detection. These findings highlight the versatility of tr-RIXS as a powerful tool for studying quantum systems far from equilibrium and their critical phenomena.
\end{abstract}
\maketitle

\section{Introduction}\label{sec:intro}

Exploring nonequilibrium dynamics in quantum systems is a rapidly advancing area of condensed matter physics, uncovering phenomena inaccessible in equilibrium. Notable examples include light-induced metal-insulator transitions~\cite{Cavalleri2001,mor2017ultrafast,okazaki2018photo, tang2020non, choi2024light}, transient superconductivity~\cite{Fausti2011,Mitrano2016,budden2021evidence,eckhardt2024theory}, the emergence of topological states~\cite{Wang2013,Mahmood2016,mogi2024direct}, and light-driven anomalous Hall effects~\cite{McIver2020}. These discoveries highlight the critical importance of nonequilibrium phases of matter and underscore the need to probe, understand, and manipulate quantum systems in regimes far from equilibrium~\cite{delaTorreReview2021}. Gaining deeper insights into these phenomena is essential for advancing our understanding of fundamental physics and exploring their potential technological applications.

To unravel the complexities of nonequilibrium dynamics, it is often instructive to begin with simple, well-understood model systems. From a theoretical perspective, the transverse field Ising chain (TFIC), known for its exact solvability and simple phase diagram~\cite{Lieb1961,Pfeuty1970}, serves as an ideal model for exploring nonequilibrium dynamics and quantum phase transitions~\cite{Dziarmaga2005,rossini2010long,Calabrese2011,Calabrese2012,essler2012dynamical,sharma2016slow,Puskarov2016,kennes2018controlling,das2025insights}. These studies often employ quench setups, where a sudden change in a Hamiltonian parameter drives the system into a nonequilibrium state. The resulting nonthermal superposition of eigenstates of the post-quench Hamiltonian evolves dynamically, giving rise to phenomena such as dynamical quantum phase transitions (DQPTs)~\cite{Heyl2013,Heyl2018}, which are marked by zeros in the Loschmidt amplitude defined as the overlap between the initial and time-evolved states. Analogous to equilibrium phase transitions but occurring in the time domain, DQPTs provide a powerful framework for understanding the interplay between criticality and dynamics in nonequilibrium quantum systems.

Experimental studies of nonequilibrium behavior have thus far been largely limited to quantum simulation platforms employing synthetic quantum materials~\cite{Meinert2013, Langen2015review, Jurcevic2017, Zhang2017, Bernien2017, Guo2019}, where precise control over parameters such as magnetic interactions is more readily achievable. Extending such investigations to condensed matter systems, which offer greater material complexity and experimental relevance, is crucial for bridging the gap between controlled theoretical models and experimentally accessible materials. In this context, TFIC serves as a limiting case for quasi-one-dimensional magnets~\cite{Yoshizawa1979, coldea2010, Bera2014, wang2016confined, Wang2018, amelin2020experimental} subjected to an external field.
However, experimental studies of the nonequilibrium dynamics in these systems remain scarce. This scarcity stems from several key challenges. First, controlling interaction parameters in condensed matter systems is significantly more difficult than in engineered quantum platforms. Second, probing their dynamics, governed by magnetic energy scales typically on the order of a few millielectronvolts, requires ultrafast techniques with both high temporal and energy resolution. Methods such as inelastic neutron scattering commonly used for probing magnetic excitations in these systems lack the requisite time resolution. Finally, many of the theoretical quantities central to characterizing nonequilibrium dynamics and DQPTs~\cite{Karrasch2013, Heyl2014, Heylet2018, piccitto2019dynamical, halimeh2021local, yu2021correlations, Haldar2020, de2021entanglement}, including the Loschmidt amplitude~\cite{Heyl2013, Heyl2018} and entanglement measures~\cite{Haldar2020, de2021entanglement}, require direct access to the quantum many-body wave function, which is inherently challenging in condensed matter experiments.

Despite these challenges, significant progress has been made toward overcoming several of the key experimental limitations discussed above. Notably, various proposals have been put forward to manipulate exchange couplings in magnetic materials using external electric or electromagnetic fields~\cite{Mentink2015, Mentink2017, Eckstein2017, Barbeau2019}, opening new avenues for dynamic control of magnetic interactions. Concurrently, there have been substantial experimental advancements in ultrafast pump-probe spectroscopic methods capable of resolving transient dynamics and disentangling the interplay between spin, charge, and orbital degrees of freedom~\cite{Fischer2016, Inoue2016, giannetti2016ultrafast}. Among them, resonant inelastic x-ray scattering (RIXS) has played a crucial role in equilibrium studies, providing precise insights into collective excitations, such as magnons, phonons, and orbital modes~\cite{Ament2011,Jia2016,Haverkort2010,Lu2017}. The emerging technique of time-resolved RIXS (tr-RIXS) offers a promising pathway for extending these capabilities to nonequilibrium regimes~\cite{Wang2017,Chen2019,mitrano2020probing,Chen2020,ejima2022nonequilibrium}. By offering real-time access to the dynamics of pump-excited quantum materials, tr-RIXS provides an unparalleled opportunity to probe and understand quantum systems far from equilibrium.

In this work, we investigate the nonequilibrium dynamics of TFIC under various pump configurations with tr-RIXS as a probe. Through a combination of equilibrium and nonequilibrium analysis, we demonstrate how tr-RIXS spectra encode information about the system's transient states and instantaneous Hamiltonian parameters. A key finding is the discovery of a one-to-one correspondence between DQPTs and oscillatory features in the low-energy tr-RIXS spectra, providing a potential experimental indicator for DQPTs. This paper is organized as follows. Section~\ref{sec:method} provides an overview of the TFIC model, the pump-probe setups, and a brief introduction to the tr-RIXS method, highlighting its connection to the time-dependent correlation functions in the TFIC. Additionally, this section introduces the concept of DQPTs and outlines their derivation in a nonequilibrium TFIC. Section~\ref{sec:results} presents a detailed discussion of the nonequilibrium spectral features, focusing on the correspondence between spectral features, instantaneous Hamiltonian parameters, and DQPTs. Section~\ref{sec:conclusion} summarizes the main findings and discusses their implications. Technical details and derivations are provided in the appendices.

\section{Methods}\label{sec:method}

\subsection{Model and setup}\label{sec:method:model}

The model we studied is the spin-1/2 TFIC
\begin{equation}\label{eq:tfic}
    H=J\sum_{i=1}^{N} \sigma_i^x \sigma_{i+1}^x - h \sum_{i=1}^{N} \sigma_i^z,
\end{equation}
where $\sigma^\alpha$ are the Pauli matrices related to the spin components $S^\alpha = \sigma^\alpha/2$ ($\alpha=x,\,z$). The nearest-neighbor spins are coupled via a magnetic exchange $J$ in the $x$-direction and subject to an external transverse magnetic field $h$ in $z$. As a paradigmatic model for studying quantum phase transitions, it has been extensively studied theoretically~\cite{Lieb1961,Pfeuty1970} and experimentally~\cite{coldea2010,Bera2014,Wang2018}. The ground-state phase diagram for TFIC exhibits a magnetically ordered phase for $|h/J|<1$ and a paramagnetic phase for $|h/J|>1$, separated by a quantum critical point at $|h/J|=1$. Depending on the exchange mechanism, $J$ can be either ferromagnetic ($J<0$) or antiferromagnetic ($J>0$). The case for $J<0$ can be mapped onto $J>0$ by a $\pi$-rotation of every second spin around the $z$-axis. Without loss of generality, we assume an antiferromagnetic coupling ($J>0$)~\cite{Bera2014,Wang2018} and $h>0$ for the subsequent discussion.

To explore the nonequilibrium dynamics of the TFIC, controlled manipulation of the model parameters $h$ and/or $J$ is crucial. In condensed matter systems, the external magnetic field $h$ can be adjusted relatively easily in experiments, whereas controlling the exchange coupling $J$ is more challenging. Magnetic couplings, being effective parameters derived from high-energy electron virtual hoppings in strongly correlated Mott insulators, require indirect manipulation. Several approaches have been proposed to modify $J$ using external electric fields or optical pulses that interact with electronic degrees of freedom~\cite{Mentink2015, Mentink2017, Eckstein2017, Barbeau2019}. These techniques have successfully demonstrated the ability to tune both the magnitude and sign of $J$. Given the routine application of magnetic, electric, and optical pump techniques in pump-probe spectroscopic experiments~\cite{Neppl2015}, we treat both $h$ and $J$ as controllable parameters in this study.

The profile of the time-varying coupling $J(t)$ depends on the specific pump mechanism and experimental setup. Generally, the amplitude of $J(t)$ approximately follows the pump’s amplitude profile~\cite{Mentink2015, Mentink2017, Eckstein2017, Barbeau2019}. For simplicity, and as a close approximation to exact results, we assume a Gaussian profile centered at the pump time $t_0$:
\begin{equation}\label{eq:jt}
J(t) = J_0 + \Delta J \cos[\omega(t-t_0)]e^{-(t-t_0)^2/2\sigma_J^2},
\end{equation}
where $\omega$ is the oscillation frequency, $\Delta J$ the maximum modulation, and $\sigma_J$ the pulse width. We will focus on few-cycle pulses where $\sigma_J \sim 2\pi/\omega$, as typically implemented in experimental setups~\cite{Fausti2011,Mitrano2016}. Similarly, we model the time-dependent magnetic field $h(t)$ as
\begin{equation}\label{eq:ht}
h(t) = h_0 + \Delta h e^{-(t-t_0)^2/2\sigma_h^2},
\end{equation}
where a static field $h_0$ is combined with a Gaussian-shaped pulsed field with strength $\Delta h$ and width $\sigma_h$.

\subsection{Time-resolved RIXS}\label{sec:method:trrixs}

To characterize the nonequilibrium dynamics and elementary excitations of the TFIC, we employ the tr-RIXS technique. In a general pump-probe setup, treating the probe pulse as a weak perturbation~\cite{Freericks2009, Chen2019,Wang2020}, the tr-RIXS cross section is defined as
\begin{equation}\label{eq:trrixs:crsec1}
\begin{split}
    I(\vb{q}, \Omega, t_\mathrm{pr}) = & \int_{\mathclap{\tau_1 \leq \tau_2, \tau_3 \leq \tau_4\mystrut}} \dd^4 \bm{\tau} \, e^{\ii \omega_i (\tau_3-\tau_1) - \ii \omega_f (\tau_4 - \tau_2)} l(\tau_1,\tau_2) l(\tau_3,\tau_4) \times \\
    & g(\tau_1,t_\mathrm{pr})g(\tau_3,t_\mathrm{pr}) \mathcal{S}_{\bm{\epsilon}_i\bm{\epsilon}_f}(\vb{q},\bm{\tau}),
\end{split}
\end{equation}
where $\Omega=\{\omega_i,\bm{\epsilon}_i,\vb{k}_i;\omega_f,\bm{\epsilon}_f,\vb{k}_f=\vb{k}_i-\vb{q}\}$ specifies the frequency, polarization, and momentum of the incident and emitted photons of the probe pulse. The term $\mathcal{S}_{\bm{\epsilon}_i\bm{\epsilon}_f}(\vb{q},\bm{\tau})$ is the Fourier transform of a four-point correlation function
\begin{equation}\label{eq:trrixs:crsec2}
\begin{split}
    \mathcal{S}_{\bm{\epsilon}_i\bm{\epsilon}_f}(\vb{q},\bm{\tau}) = & \sum_{m,n} e^{\ii \vb{q}\cdot(\vb{r}_m-\vb{r}_n)} \times \\
    &\langle D^\dag_{\bm{\epsilon}_i n}(\tau_3) D_{\bm{\epsilon}_f n}(\tau_4) D^\dag_{\bm{\epsilon}_f m}(\tau_2) D_{\bm{\epsilon}_i m}(\tau_1) \rangle,
\end{split}
\end{equation}
where the dipole operators $D_{\bm{\epsilon} m} = \sum_{\mu\nu}M_{\bm{\epsilon}}^{\mu \nu} d^\dag_{\mu} c_{\nu}$ describe transitions at site $m$ from core states with spin-orbital index $\nu$ to valence states $\mu$ at resonance, with dipole matrix elements $M_{\bm{\epsilon}}^{\mu \nu}$ depending on the light polarization~\cite{deGrootBook}. In the Heisenberg picture, it is given by $D_{\bm{\epsilon} m}(\tau) = U(-\infty, \tau) D_{\bm{\epsilon} m} U(\tau, -\infty)$, where the time-evolution operator $U(\tau, -\infty) = \mathcal{T} \exp[-\ii \int_{-\infty}^{\tau}\dd t H(t)]$ is defined with respect to a time-dependent Hamiltonian considering the external pumps in Sec.~\ref{sec:method:model}. Throughout the remainder of the paper, zero temperature is assumed, and all expectation values $\langle \cdot \rangle$ are taken with respect to the ground state. The functions $g(\tau_i,t_\mathrm{pr})$ are the probe pulse profiles associated with the incident photons, assumed to take a Gaussian form
\begin{equation}\label{eq:probe}
    g(\tau_i,t_\mathrm{pr})=\frac{1}{\sqrt{2\pi}\sigma_\mathrm{pr}}\exp [-(\tau_i-t_\mathrm{pr})^2/(2\sigma_\mathrm{pr}^2)],
\end{equation}
centered at the probe time $t_\mathrm{pr}$ with width $\sigma_\mathrm{pr}$. A function $l(\tau_1, \tau_2)$ that decays with increasing time difference $|\tau_1-\tau_2|$ is introduced to account for decay processes of the intermediate excited states due to degrees of freedom not included in the model.

It is evident from Eq.~\eqref{eq:trrixs:crsec2} that tr-RIXS probes various correlation functions of the valence electrons of the form $\langle d_{n\kappa}(\tau_3) d^\dag_{n\lambda}(\tau_4) d_{n\nu}(\tau_2) d^\dag_{n\mu}(\tau_1)\rangle$. In equilibrium, both qualitative and quantitative~\cite{Ament2007,Jia2016,Haverkort2010,Lu2017} relations have been established between RIXS cross sections and different spin, charge, or current correlations. Notably, for magnetic Mott insulators, these relations show that at low energies, the RIXS cross section is dominated by spin correlation functions when the incident and emitted photons have crossed polarizations, as other low-energy excitations are gapped out.

To derive an approximate relation between the tr-RIXS cross section~\eqref{eq:trrixs:crsec1} and the intrinsic correlation functions of the system, we assume an ultra-short core-hole lifetime by setting $l(\tau_1, \tau_2) = \delta(\tau_1, \tau_2)$. This assumption corresponds to the instantaneous relaxation of the core hole after its creation and is justified because the core hole is highly unstable, with a (sub)femtosecond lifetime~\cite{deGrootBook}. This timescale corresponds to dynamics at energy scales of electron volts, whereas the low-energy excitations of interest, such as magnetic excitations in the spin Hamiltonian, are typically on the order of millielectronvolts. Under this assumption, for a spin-1/2 system, Eq.~\eqref{eq:trrixs:crsec2} can be decomposed into two-point correlation functions of operators $d_{n\nu}d^\dag_{n\mu}(\tau)$, which are either of charge type ($\mu=\nu\in \{\uparrow, \downarrow\}$) or spin type ($\mu \neq \nu$) depending on the light polarization. In the cross-polarized channel, such an approximation was also shown previously to reproduce the exact RIXS cross section up to an overall scaling factor~\cite{Jia2016}. Consequently, in the cross-polarization channel, the low-energy tr-RIXS cross section can be expressed as a weighted spin-spin correlation function
\begin{equation}\label{eq:trrixs:crsec3}
\begin{split}
    I(\vb{q}, \omega, t_\mathrm{pr}) = & \int_{-\infty}^{\infty} \dd^2 \bm{\tau} \, e^{\ii \omega (\tau_2-\tau_1)} \times \\
    & g(\tau_1, t_\mathrm{pr}) g(\tau_2, t_\mathrm{pr}) S^{\alpha\alpha}(\vb{q},\tau_1,\tau_2),
\end{split}
\end{equation}
where $\omega=\omega_i-\omega_f$ is the energy loss. An overall scaling factor dependent on the incident frequency and light polarizations is omitted. The spin-spin correlation function is defined as $S^{\alpha\alpha}(\vb{q},\tau_1,\tau_2)=1/N \sum_{m,n} e^{\ii \vb{q}(r_m-r_n)}\ev{S^\alpha_m(\tau_2)S^\alpha_n(\tau_1)}$, where $N$ is the system size. Equation~\eqref{eq:trrixs:crsec3} generalizes similar expressions derived for the equilibrium~\cite{Haverkort2010}, which is recovered by setting the probe profile to a constant, $g(\tau, t_\mathrm{pr})=1$. In such a case, Eq.~\eqref{eq:trrixs:crsec3} reduces to the spin dynamical structural factor (DSF) $S^{\alpha\alpha}(\vb{q},\omega) = \int_{-\infty}^{\infty} \dd \tau \, e^{\ii \omega \tau} S^{\alpha\alpha}(\vb{q},0,\tau)$.

The energy-integrated RIXS cross section, corresponding to standard resonant X-ray scattering (tr-RXS), is another widely used technique for probing lattice and electronic structures. Integrating Eq.~\eqref{eq:trrixs:crsec3} over $\omega$ yields \begin{equation}\label{eq:trrixs:trrxs}
    I_0(\vb{q}, t_\mathrm{pr}) = \int_{-\infty}^{\infty} \dd \tau g(\tau,t_\mathrm{pr})^2 S^{\alpha\alpha}(\vb{q},\tau,\tau),
\end{equation}
which, in the short-time probe limit ($\sigma_\mathrm{pr} \ll 1$) measures the equal-time spin structure factor at the probe time.

\subsection{Correlation functions of TFIC}\label{sec:method:corfun}

After establishing the relation between the tr-RIXS cross section and the dynamical as well as the equal-time correlation functions, we proceed to derive these quantities for the TFIC.

In equilibrium, the TFIC Hamiltonian~\eqref{eq:tfic} with periodic boundary conditions can be solved by mapping it to a model of spinless fermions via the Jordan-Wigner transformation~\cite{Lieb1961}, where the spin operators are expressed in terms of the fermion operators as $\sigma_i^z=1-2c^\dag_i c_i$ and $\sigma_i^x=\Pi_{l=1}^{i-1}(1-2c_l^\dag c_l)(c_i+c_i^\dag)$. This mapping transforms Eq.~\eqref{eq:tfic} into a quadratic form in fermionic operators, which can then be diagonalized using a Bogoliubov transformation in momentum space
\begin{equation*}
    c_{k}=u_{k}\eta_{k}+\mathrm{i} v_{k} \eta_{-k}^{\dagger}.
\end{equation*}
The operators $c_{k}=1/\sqrt{N} \sum_{i=1}^N e^{\ii k i} a_{i}$ are defined for momenta $k=2\pi(n+1/2)/N$, where $n=-\frac{N}{2},...,\frac{N}{2}-1$, in the Neveu-Schwarz sector with an even number of fermions. Required by the fermion anticommutation relations, the Bogoliubov coefficients $u_k=\cos \frac{\theta_k}{2}$ and $v_k=\sin \frac{\theta_k}{2}$ satisfy $(u_k)^2 + (v_k)^2 = 1$ as well as $u_k=u_{-k}$ and $v_k=-v_{-k}$. Here, $\theta_k$ is determined by
\begin{equation}
    \tan \theta_k = \frac{ J\sin k}{h+J\cos k},
\end{equation}
assuming $J, h>0$. The diagonalized Hamiltonian becomes $H=\sum_k \epsilon_k(\eta^\dag_k \eta_k - \frac{1}{2})$, whose eigen spectrum is given as
\begin{equation}\label{eq:ek}
    \epsilon_k=2\sqrt{J^2+h^2+2Jh \cos k },
\end{equation}
with an excitation gap at $k=\pi$ given by $\Delta = 2|J-h|$, which vanishes at the quantum critical point $h/J=1$.

Under an external pump, the Hamiltonian $H(t)$ becomes time-dependent due to the temporal variation of the parameters $J(t)$ and/or $h(t)$. The Bogoliubov transformation thus evolves in time as~\cite{Dziarmaga2005, Puskarov2016}
\begin{equation}
    c_{k}(t)=\mu_{k}(t)\eta_{k}+\mathrm{i} \nu_{k}(t) \eta_{-k}^{\dagger},
\end{equation}
where the coefficients satisfy the equations of motion
\begin{equation}
\label{eq:time bogo coeff}
\ii \frac{\dd}{\dd t}{ \mu_k(t)\brack \nu_{-k}^*(t)}=\left [\begin{array}{cc}
A_k(t) & B_k(t) \\
B_k(t) & -A_k(t)
\end{array} \right ]{ \mu_k(t)\brack \nu_{-k}^*(t)}
\end{equation}
with $A_k(t)=2h(t)+2J(t)\cos k$ and $B_k(t)=-2J(t)\sin k$.

In this work, we focus on the transverse spin-spin correlation function
\begin{equation}\label{eq:ss}
\begin{split}
    S^{zz}(q,\tau_1,\tau_2) &= \frac{1}{N} \sum_{k,k'}\left \langle c_k^+(\tau_2)c_{k+q}(\tau_2)c_{k'}^+(\tau_1)c_{k'-q}(\tau_1) \right \rangle   \\
 &=\frac{1}{N} \sum_{k}  [\mu_{k+q}(\tau_2)\mu^*_{k+q}(\tau_1)\nu_{k}(\tau_1)\nu^*_{k}(\tau_2) + \\
&\quad \quad \mu_{k+q}(\tau_2)\mu^*_{k}(\tau_1)\nu_{k+q}(\tau_1)\nu^*_{k}(\tau_2)].
\end{split}
\end{equation}
For a general pump form $J(t)$ or $h(t)$, the coefficients $\mu_k(t)$ and $\nu_k(t)$ do not admit analytic expressions and are solved numerically for different pump configurations. This expression can then be used to compute the tr-RIXS or tr-RXS intensities using Eq.~\eqref{eq:trrixs:crsec3} or Eq.~\eqref{eq:trrixs:trrxs}, respectively.

\subsection{Dynamical quantum phase transitions in TFIC}\label{sec:method:dqpt}

To investigate correlations between time-resolved spectroscopic quantities and changes in the system’s physical properties, we compute the Loschmidt amplitude, which quantifies the overlap of the time-evolved state $|\psi(t)\rangle$ and the initial ground state $| \psi^i \rangle$~\cite{Heyl2018}
\begin{equation}\label{eq:loschmidt}
    G(t) = \langle \psi^i  | \psi(t)  \rangle.
\end{equation}
The Loschmidt rate function, defined as
\begin{equation}\label{eq:lt}
    l(t)=-\frac{1}{N}\mathrm{ln} |G(t)|^2,
\end{equation}
provides a measure of DQPTs, identified by nonanalytic structures in $l(t)$. Calculating these quantities requires the evaluation of $| \psi(t) \rangle = U(t) | \psi^i\rangle$.

In the Bogoliubov fermion space, the time-dependent Hamiltonian reads
\begin{equation}\label{eq:trH}
\begin{split}
    H(t)=&\sum_{k>0} 2\{A_k(t)[(u_k^i)^2-(v_k^i)^2]-2B_k(t)u_k^iv_k^i\}K_k^0 \\
&+i \{2A_k(t)u_k^iv_k^i+B_k(t)[(u_k^i)^2-(v_k^i)^2]\} \\
& \times (K_k^++K_k^-),
\end{split}
\end{equation}
where $u_k^i$, $v_k^i$ are the initial Bogoliubov coefficients that diagonalize the pre-pump Hamiltonian, and
\begin{equation}
\begin{split}
    K_k^0=&\frac{1}{2}(\eta_{k}^+\eta_{k}+\eta_{-k}^+\eta_{-k}),\\
    K_k^+=&\eta_{k}^+\eta^+_{-k}, \\
    K_k^-=&\eta_{k}\eta_{-k}.
\end{split}
\end{equation}
Since Eq.~\eqref{eq:trH} only couples pairs of momenta $k$ and $-k$, the time-evolution operator factorizes as $U(t) = \prod_{k>0} U_k(t)$, allowing $G(t)$ and $l(t)$ to be evaluated mode by mode. Given that the terms $K_k^0$, $K_k^+$, and $K_k^-$ satisfy the $SU(1,1)$ Lie algebra $[ K_k^-, K_p^+]$$ =2\delta_{kp}K_k^0$ and $[ K_k^0, K_p^\pm ]$$=\pm\delta_{kp}K_k^{\pm}$, one could make the ansatz for the time-evolution operators ~\cite{truax1985baker,dora2013loschmidt,Puskarov2016}
\begin{equation}\label{eq:Uk}
U_k(t)=e^{i\phi_k(t)}e^{C_k^+(t)K_k^+}e^{C_k^0(t)K_k^0}e^{C_k^-(t)K_k^-}.
\end{equation}
Substituting Eq.~\eqref{eq:Uk} into the Schr\"{o}dinger equation of the time evolution operator, $i\dot{U}_k(t)=H_k(t)U_k(t)$, yields the expressions
\begin{eqnarray}
\begin{split}
\phi_k(t)=&i\int_0^t \mathrm{d}t' \{2A_k(t')u_k^iv_k^i+B_k(t')[(u_k^i)^2-(v_k^i)^2]\}, \\
C_k^0(t)=&-2\ln{[u_k^i\mu_k^*(t)+v_k^i \nu_k(t)]}, \\
C_k^+(t)=&\frac{-i[v_k^i\mu_k^*(t)-u_k^i\nu_k(t)]}{u_k^i\mu_k^*(t)+v_k^i \nu_k(t)},\\
C_k^-(t)=&\frac{i[v_k^i\mu_k(t)-u_k^i \nu^*_k(t)]}{u_k^i\mu_k^*(t)+v_k^i \nu_k(t)},
\end{split}
\end{eqnarray}
where $\mu_k(t),\nu_k(t)$ are the time-dependent Bogoliubov coefficients in Eq.~\eqref{eq:time bogo coeff}. Applying Eq.~\eqref{eq:Uk} on the initial state $| \psi_k^i \rangle=$$[u_k^i , iv_k^i ]^T$ yields the time-dependent wave function $| \psi_k(t) \rangle=U_k(t) | \psi_k^i \rangle=e^{i\phi_k(t)}[\mu^*_{k}(t),i\nu_{k}(t)]^T$. The Loschmidt rate function is then given as
\begin{eqnarray}
\begin{split}
l(t)=&-\frac{1}{\pi} \int_{0}^{\pi} \mathrm{d}k\ \mathrm{ln}  |  \langle \psi_k^i  | \psi_k(t)   \rangle   |\\
=&-\frac{1}{\pi} \int_{0}^{\pi} \mathrm{d}k\ \mathrm{ln}  | u_k^i \mu_k^*(t)+v_k^i \nu_k(t)  |.
\label{eq:ltTFIC}
\end{split}
\end{eqnarray}

\begin{figure}[tb]
    \centering
    \includegraphics[width=\linewidth]{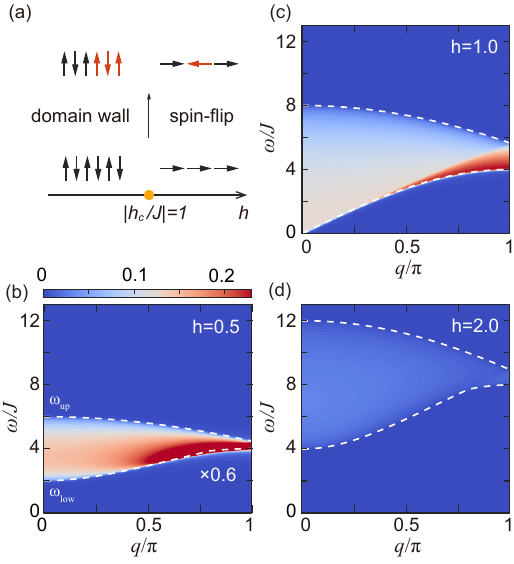}
    \caption{\label{fig:eq} (a) Phase diagram of the TFIC and the corresponding quasiparticles in the ordered and disordered phases. (b)-(d) Equilibrium RIXS, or transverse DSF $S^{zz}(q,\omega)$ for the TFIC with $J=1.0$ and $h=0.5$, $1.0$, and $2.0$, respectively. The white dashed lines mark the boundaries of the two-particle continua. The intensity in (b) is scaled by $0.6$. }
\end{figure}

\begin{figure*}[t]
  \centering
  \includegraphics[width=\textwidth]{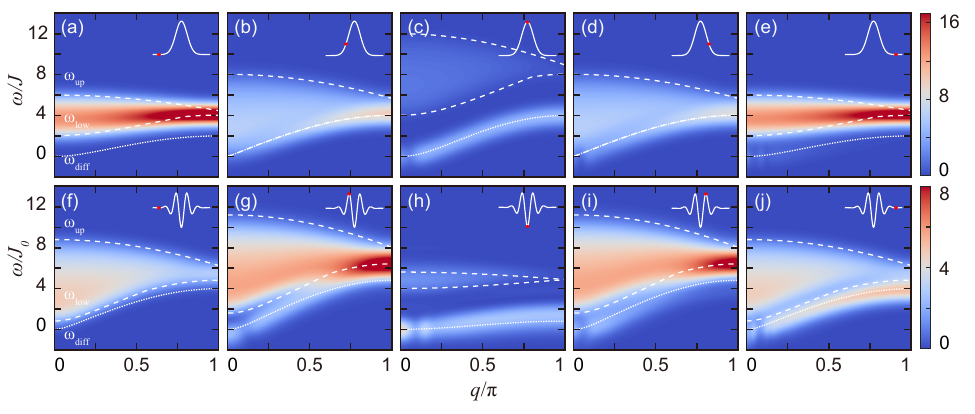}
  \caption{\label{fig:noneq} (a)-(e) Nonequilibrium tr-RIXS spectra during an $h$ pump with parameters $h_0=0.5$, $\Delta h=1.5$, and $\sigma_h=4$, with fixed $J=1.0$. The measurement times are $t_\mathrm{pr}=$ $-15.6$, $-5.9$, $0.0$, $5.9$, and $15.6$ relative to the pump time $t_0=0$, as indicated by the red dots in the sketches in each panel, corresponding to $h(t) =$ $0.5$, $1.0$, $2.0$, $1.0$, and $0.5$, respectively. (f)-(i) Nonequilibrium tr-RIXS spectra during a $J$ pump with parameters $J_0=1.0$, $\Delta J=-0.8$, $\omega=0.8$, and $\sigma_J=5$, with constant $h=1.2$. The measurement times are at $t_\mathrm{pr}=$ $-15.6$, $-3.7$, $0.0$, $3.7$, $15.6$, corresponding to $J(t)=1.0$, $1.6$, $0.2$, $1.6$, and $1.0$, respectively. All spectra are measured with $\sigma_\mathrm{pr}=1.0$. In each panel, the upper two dashed lines indicate the boundaries of the instantaneous two-kink continua determined by $H(t_\mathrm{pr})$, while the lowest dotted line represents the energy difference $\omega_\mathrm{diff}(t_\mathrm{pr}) = \epsilon_{\pi+q}(t_\mathrm{pr})-\epsilon_\pi(t_\mathrm{pr})$.}
\end{figure*}

\section{Results}\label{sec:results}

\subsection{Equilibrium and nonequilibrium Tr-RIXS spectra}
Figure~\ref{fig:eq} shows the equilibrium RIXS cross section, or the transverse DSF $S^{zz}(q,\omega)$ of the TFIC across the quantum critical point at $h/J=1$. These results serve as a reference for the nonequilibrium discussion presented later. All calculations in this work used a chain length of $N=300$. We verified that the conclusions are insensitive to system size. As shown in Fig.~\ref{fig:eq}(a), the elementary excitations in the TFIC are domain walls (kinks) separating the two degenerate antiferromagnetically ordered ground states in the $x$ direction, each with an energy of $2J$ at $h=0$. A non-zero transverse field $h$ induces hopping between these kinks via the operators $\sigma^z_i$, allowing the single-kink states to acquire dispersions $\epsilon_k$ in momentum space, as defined by Eq.~\eqref{eq:ek}.

The transverse DSF, which measures the two-kink states, exhibits non-zero weight in the energy-momentum space $(\omega,q)$ that satisfies $\omega=\epsilon_{k_1} + \epsilon_{k_2}$ and $q=k_1+k_2$ for all $k_1,k_2$ in the Brillouin zone. The upper and lower boundaries of these continua are defined by $\omega_{\text{up}} = 2\epsilon_{q/2}$ and $\omega_{\text{low}} = \min(2\epsilon_{\pi+q/2}, \epsilon_{\pi} + \epsilon_{\pi+q})$, respectively, shown as dashed lines in Figs.~\ref{fig:eq}(b)-(d). In the small-field limit, DSF intensity is concentrated within a narrow energy range around $4J$, as shown in Fig.~\ref{fig:eq}(b). As $h$ increases, the energy distribution broadens while the overall intensity of $S^{zz}(q,\omega)$ decreases [Fig.~\ref{fig:eq}(c)], reflecting the external field’s disruption of magnetic ordering and destruction of domain walls. In the large-field limit, the elementary excitations reduce to incoherent local spin flips in the $z$-direction with vanishing correlations, as reflected by the negligible intensities in Fig.~\ref{fig:eq}(d).

We now turn to the nonequilibrium case, considering pumps that modulate the coupling $J(t)$ or field $h(t)$ as defined in Eqs.~\eqref{eq:jt} and \eqref{eq:ht}. Figure~\ref{fig:noneq} shows the nonequilibrium tr-RIXS $I(q, \omega, t_\mathrm{pr})$ [Eq.~\eqref{eq:trrixs:crsec3}] measured at different probe times $t_\mathrm{pr}$ during a $J$ or $h$ pump. During an $h$ pump in which $J$ is held constant and $h(t)/J$ varies from 0.5 to 2.0 and back to 0.5, the system crosses the equilibrium quantum critical point twice. The tr-RIXS spectra exhibit features reminiscent of the equilibrium case. At the initial stage of the pump [Fig.~\ref{fig:noneq}(a)], where the change in $h(t)$ is negligible, the spectrum displays weight distributed approximately within the boundaries of the two-kink continuum determined by the static parameters at $h/J=0.5$. Compared to its equilibrium counterpart in Fig.~\ref{fig:eq}(b), the spectrum is broadened due to the finite probe width ($\sigma_\mathrm{pr} = 1.0$) and the loss of time-translation symmetry. As $h(t)$ increases, the similarity between the equilibrium and nonequilibrium spectra remains apparent, with both exhibiting a vanishing gap at the equilibrium quantum critical point $h/J=1$ in Fig.~\ref{fig:eq}(c) and Fig.~\ref{fig:noneq}(b).

Significant differences arise when $h(t)$ exceeds $J$, as seen in Fig.~\ref{fig:noneq}(c). While the spectrum still exhibits the two-kink continuum, with boundaries determined by substituting the instantaneous field at the probe time $h(t_\mathrm{pr})=2.0$ into Eq.~\eqref{eq:ek}, similar to Fig.~\ref{fig:eq}(d), an additional low-energy feature appears within the original spectral gap. This reflects the presence of a distinct two-particle correlation in the pump-excited system. In nonequilibrium, the time-dependent state $|\psi_k(t)\rangle$ is no longer an eigenstate of the instantaneous Hamiltonian $H(t)$ but instead becomes a linear combination of the eigenstates, $|\psi_k(t)\rangle=$ $\alpha_k^+(t) |\psi^+_k(t)\rangle + \alpha_k^-(t) |\psi^-_k(t)\rangle$, where $|\psi^+_k(t)\rangle$ and $|\psi^-_k(t)\rangle$ correspond to the eigenenergies $\epsilon_k(t)$ and $-\epsilon_k(t)$, respectively. Its time evolution follows $|\psi_k(t+\delta t)\rangle \sim \alpha_k^+(t) e^{-i\epsilon_k(t)\delta t} |\psi^+_k(t)\rangle + \alpha_k^-(t) e^{i\epsilon_k(t)\delta t} |\psi^-_k(t)\rangle$, a superposition of two Bogoliubov waves with opposite frequencies. Consequently, the two-particle excitations include contributions from terms involving quasiparticles with negative energies. In addition to the two-particle continuum observed in equilibrium, this leads to additional continua within the energy windows such as $\omega=\epsilon_{k+q} - \epsilon_{k}$. However, the intensity distribution of the additional spectral feature in Fig.~\ref{fig:noneq}(c) follows approximately a well-defined dispersion defined by $\omega_\mathrm{diff}(t_\mathrm{pr}) = \epsilon_{\pi+q}(t_\mathrm{pr})-\epsilon_\pi(t_\mathrm{pr})$ rather than forming a continuum. This can be understood as, in the excited states, single-particle excitations are most easily generated at momenta corresponding to the smallest gap at  $k=\pi$ in Eq.~\eqref{eq:ek}. This also explains the apparent absence of this feature in Figs.\ref{fig:noneq}(b) and (d), where $\omega_\mathrm{diff}$ coincides with $\omega_\mathrm{low}$. Additionally, the tr-RIXS cross section contains other excitations that are less observable in most settings. A complete analysis of these is provided in App.~\ref{app:crosssection}. After the $h$ pump, the tr-RIXS spectrum in Fig.~\ref{fig:noneq}(e) recovers to a state similar to the pre-pump spectrum in Fig.~\ref{fig:noneq}(a), with the additional feature exhibiting negligible intensity.

The evolution of the tr-RIXS during a $J$ pump with constant $h$, shown in Fig.~\ref{fig:noneq}(f)-(j), follows a pattern similar to that of the $h$ pump. As the pump progresses, the near-equilibrium pre-pump spectrum in Fig.~\ref{fig:noneq}(f) gradually develops the additional energy-difference excitation at $\omega_\mathrm{diff}$ in Fig.~\ref{fig:noneq}(g), with its intensity peaking at the height of the pump in Fig.~\ref{fig:noneq}(h) and diminishing again in Fig.~\ref{fig:noneq}(i). Notably, unlike the $h$ pump considered above, where the post-pump spectrum nearly returns to the pre-pump state, the spectrum here remains distinct, featuring a significant $\omega_\mathrm{diff}$-excitation, indicating that the system remains in a highly excited states.

\begin{figure}[tb]
    \centering
    \includegraphics[width=\linewidth]{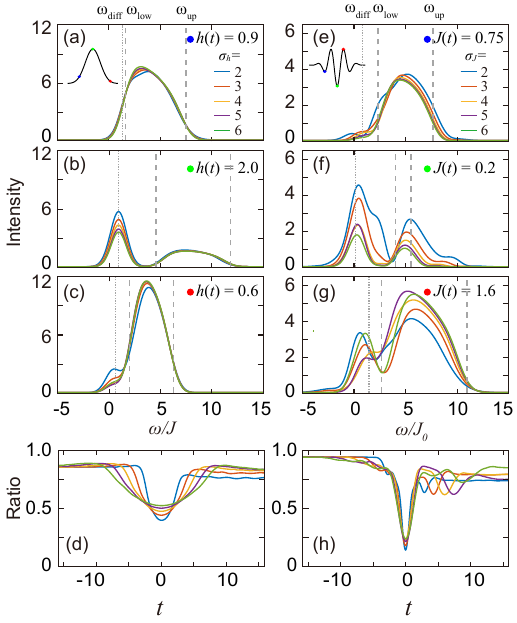}
    \caption{\label{fig:configs} (a)-(c) Tr-RIXS spectra at $q = \pi/4$ measured at various probe times $t_\mathrm{pr}$ during an $h$-pump with $h_0 = 0.5$, $\Delta h = 1.5$, and varying $\sigma_h$, with $J = 1.0$ held constant. (d) Time evolution of the spectral weight ratio within the instantaneous two-particle continuum, bounded by $\omega_\mathrm{low}$ and $\omega_\mathrm{up}$. (e)-(h) Corresponding results for a $J$-pump with $J_0 = 1.0$, $\Delta J = -0.8$, fixed $h = 1.2$, and varying $\sigma_J$, subject to $\omega_J \sigma_J = 4$. All spectra are computed with $\sigma_\mathrm{pr} = 1.0$. Insets in (a) and (e) indicate the selected probe times $t_\mathrm{pr}$. Dotted lines mark the instantaneous $\omega_\mathrm{diff}$, while dashed lines mark the instantaneous continuum boundaries $\omega_\mathrm{low}$ and $\omega_\mathrm{up}$ at $q = \pi/4$.
}
\end{figure}

To gain deeper insight into the tr-RIXS spectral features and their dependence on pump configurations, we compute spectra for various $h(t)$ and $J(t)$ scenarios. Figure~\ref{fig:configs} compares the resulting spectra for $h$ or $J$ pumps with varying widths and/or frequencies. During the $h$ pump, the system begins in a magnetically ordered state with $h_0/J = 0.5$. The $h$ pump is applied with the same amplitude but different widths $\sigma_h$. In the initial stage of the pump, where $h(t)/J < 1$ remains below the critical value, all tr-RIXS spectra exhibit similar lineshapes, with their dominant spectral weight largely confined within the two-kink continuum defined by the instantaneous boundaries $\omega_\mathrm{up}$ and $\omega_\mathrm{low}$ in Fig.~\ref{fig:configs}(a). At the peak of the pump, although the value of $h(t)$ is the same across all cases, the spectra exhibit notable differences, as shown in Fig.~\ref{fig:configs}(b). The high-energy continuum remains largely unchanged, indicating that it is primarily determined by the instantaneous Hamiltonian $H(t)$. In contrast, the $\omega_\mathrm{diff}$ excitation shows an increasing spectral weight with faster pumps, reflecting a more intense excitation of the system under rapid driving. This trend persists even after the pump subsides in Fig.~\ref{fig:configs}(c), where the most pronounced $\omega_\mathrm{diff}$ excitation is observed for the fastest pump with $\sigma_h=2$. The time evolution of the spectral weight ratio within the instantaneous continuum boundaries is shown in Fig.~\ref{fig:configs}(d). In the adiabatic limit, this quantity is expected to remain close to 1. Before the pump, the ratio is slightly below 1, which is attributed to broadening effects introduced by the finite probe width. During and after the pump, faster driving leads to a more pronounced reduction in the relative spectral weight, indicating stronger deviations from adiabatic behavior. During the $J$ pump, the system initially resides in a paramagnetic state with $h/J_0 = 1.2$. As shown in Fig.~\ref{fig:configs}(e), in the early stage of the pump where $J(t)$ undergoes only small modifications, the spectra maintain a consistent lineshape. However, for the fastest pump with $(\sigma_J, \omega_J) = (2, 2)$, a preliminary weight splitting toward lower energies becomes apparent. At the extremum of the pump in Fig.~\ref{fig:configs}(f), similar to the case of $h$ pumps, the spectral weight of the $\omega_\mathrm{diff}$ excitation increases with the pump speed. However, unlike Fig.~\ref{fig:configs}(b), the higher-energy continuum is not well described by the instantaneous boundaries, particularly for faster pumps. This discrepancy arises from the finite probe width $\sigma_\mathrm{pr} = 1$, which averages over a range of $J(t)$ values, leading to deviations for rapid pump configurations. In stark contrast to the earlier observations, the spectra in Fig.~\ref{fig:configs}(g) deviate significantly. While the spectra still exhibit a continuum defined by the instantaneous boundaries and an  $\omega_\mathrm{diff}$ excitation, the latter shows no clear trend of dependence on the pump configurations. This behavior is likely a consequence of the nontrivial time evolution of the wavefunction during the $J$ pump, where repeated crossings of the equilibrium critical point introduce complexities into the excitation dynamics. This complexity is also reflected in the spectral weight ratio shown in Fig.~\ref{fig:configs}(h), which exhibits a more intricate post-pump recovery compared to the $h$-pump case.

\subsection{Tr-RIXS and DQPTs}

\begin{figure}[tb]
    \centering
    \includegraphics[width=\linewidth]{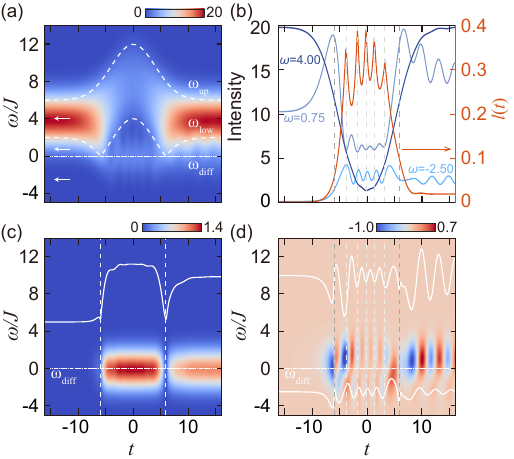}
    \caption{\label{fig:S_omega_t} (a) Time evolution of the nonequilibrium tr-RIXS spectra at $q = 0$ during the $h$ pump corresponding to Figs.~\ref{fig:noneq}(a)-(e). All spectra are measured with $\sigma_\mathrm{pr}=0.5$. The white dashed lines indicate the instantaneous two-kink continuum boundaries. (b) Constant energy cuts of the tr-RIXS spectra at energies $\omega/J =$ $4.00$, $0.75$, and $-2.50$, as indicated by white arrows in (a). The Loschmidt rate function $l(t)$ is shown for comparison, with the DQPTs marked by light gray lines. The dark gray lines indicate the time when $h(t)/J = 1$. (c)-(d) Time evolution of the intensity of (c) the  $\omega_\mathrm{diff}$  excitation ($I_{+-}$) and (d) the interference term ($I_\mathrm{int}$). The white solid lines show the energy-integrated intensity. For $I_\mathrm{int}$, the integration is performed separately for positive and negative energies. The white dotted line in panels (a), (c), and (d) represents the energy difference $\omega_\mathrm{diff} = \epsilon_{\pi+q}(t_\mathrm{pr})-\epsilon_\pi(t_\mathrm{pr})$.}
\end{figure}

\begin{figure}[tb]
    \centering
    \includegraphics[width=\linewidth]{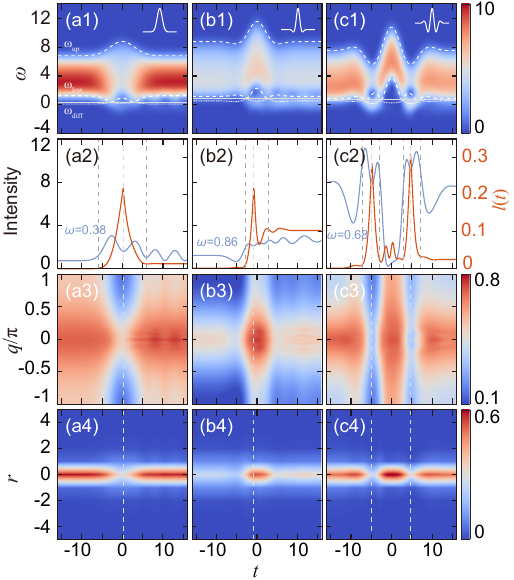}\caption{\label{fig:S_r_t} (a1)-(c1) Time evolution of the nonequilibrium tr-RIXS spectra at $q = \frac{\pi}{16}$ during the $h$ pump with parameters $h_0=0.7$, $\Delta h=0.5$, $\sigma_h=3$, and fixed $J=1.0$ (a1); at $q = \frac{\pi}{8}$ during the $J$ pump with parameters $J_0=1.0$, $\Delta J=0.7$, $\omega=0.4$, $\sigma_J=3$, and constant $h=1.2$ (b1); and at $q = \frac{\pi}{8}$ during the $J$ pump with parameters $J_0=1.0$, $\Delta J=0.8$, $\omega=0.6$, $\sigma_J=5$, and constant $h=0.9$ (c1). The dashed and dotted lines denote the continuum boundaries $\omega_\mathrm{up/low}$ and $\omega_\mathrm{diff}$, respectively. Panels below display, for each pump configuration, the constant energy cuts of the tr-RIXS spectra (indicated by light gray solid lines in panels (a1) to (c1)) over time compared to $l(t)$ [(a2)-(c2)], the energy-integrated tr-RIXS spectra in momentum space [(a3)-(c3)], and their Fourier transformations into real space [(a4)-(c4)], respectively. Light gray dashed lines denote the locations of DQPTs, while dark gray dashed lines indicate the times when $h(t)/J = 1$ or $h/J(t) = 1$.}
\end{figure}

To further investigate the relationship between the spectral features and the underlying system dynamics, we analyze their potential correlations by jointly examining the tr-RIXS spectra and the system’s evolution as characterized by the Loschmidt amplitude and rate function, throughout the duration of the pumps.

Figure~\ref{fig:S_omega_t}(a) shows the time evolution of the tr-RIXS spectra at $q = 0$ during the $h$ pump with configuration corresponding to Figs.~\ref{fig:noneq}(a)-(e). Starting in the ordered phase with $h_0/J = 0.5$, the spectra prominently feature a two-kink continuum in the initial pump stage. The spectral weight diminishes at $t_\mathrm{pr} \approx -5.9$, signaling the system’s crossing of the equilibrium critical point and the associated reconfiguration of single-particle excitations into incoherent spin flips. As the parameters cross back over the critical point into the ordered regime, the spectral weight reappears. Thus, despite the system’s nonequilibrium nature, the portion of the spectra corresponding to the two-kink continuum, analogous to its equilibrium counterpart, serves as a direct indicator of the system’s instantaneous parameters.

More intriguingly, beyond the two-kink continuum at higher energies, the low-energy spectra exhibit nontrivial oscillatory behavior over time, as highlighted by the constant energy cuts at $\omega/J =$ $0.75$ and $-2.50$ in Fig.~\ref{fig:S_omega_t}(b). A direct comparison with the Loschmidt rate function $l(t)$ reveals a striking one-to-one correspondence between the nonanalytic points of $l(t)$, indicative of DQPTs, and the minima or maxima of the spectral weight, depending on the energy of the cuts. To gain further insight into this unexpected correlation, we analyze the individual components of the tr-RIXS cross section in App.~\ref{app:crosssection}. Similar to the two-kink continuum, most components exhibit abrupt changes when the parameters cross the equilibrium critical point, as exemplified by the $\omega_\mathrm{diff}$ excitation ($I_{+-}$) shown in Fig.~\ref{fig:S_omega_t}(c). Notably, at $q = 0$, $\omega_\mathrm{diff}$ aligns with zero energy loss, indicating that such changes are also detectable with resonant elastic X-ray scattering. However, as illustrated in Fig.~\ref{fig:S_omega_t}(d), the term $I_\mathrm{int}$, arising from interference between different quasiparticle channels, uniquely exhibits an oscillatory behavior. At $q = 0$, as detailed in App.~\ref{app:crosssection}, $I_\mathrm{int}$ comprises two contributions with opposite intensities, centered around $\omega = \pm \omega_\pi$. For the positive energy contribution, its time evolution closely follows that of $l(t)$, with intensity dips in the time range $-5.9 < t < 5.9$ exhibiting a clear one-to-one correspondence with the nonanalytic points in $l(t)$. In contrast, the negative energy contribution, which displays a comparably weaker intensity, shows an opposite correspondence, with intensity peaks aligning with the nonanalytic points. As discussed in App.~\ref{app:crosssection}, these intensities are modulated by the real part of a time-dependent coefficient $g_{3,k}$, derived from the time-evolved Bogoliubov coefficients, which underpins this correspondence. A detailed derivation and elaboration of this relationship are provided in App.~\ref{app:correspondence}.
While this correspondence is mathematically well established there, we provide a more intuitive explanation here, as DQPTs, defined via the Loschmidt amplitude at specific momenta, do not obviously manifest in a collective observable like tr-RIXS. The DQPT-sensitive component $I_\mathrm{int}$ arises primarily from interference between excitation modes near $k = \pi$, where the energy gap is minimal and the system is most susceptible to nonadiabatic dynamics. As a result, although RIXS is momentum-integrated, the dominant contribution to $I_\mathrm{int}$ originates from a narrow region around $k = \pi$. The connection to the DQPT thus becomes natural, as the largest deviations between the time-evolved and initial states, which drive the DQPT [see Eq.~\eqref{eq:ltTFIC}], also occur near $k = \pi$.
We note that the low-energy spectral weight also exhibits intensity modulations after the pump ($t > 5.9$). However, their connection to the DQPTs is ruled out by the fact that DQPTs in the TFIC occur only when the system, starting from the ground state, is pumped across the equilibrium quantum critical point~\cite{Karrasch2013,Heyl2014}. Exceptions to this rule exist when the system is initially prepared in states with specific geometric properties~\cite{Hickey2014}. Therefore, a joint analysis of both the high-energy spectral weight, which reflects the instantaneous parameters relative to the equilibrium quantum critical point, and the low-energy oscillations is necessary to establish their correspondence to the DQPTs.

To investigate the general applicability of this correspondence, we examine its validity in Fig.~\ref{fig:S_r_t} across a range of $h$ and $J$ pump configurations, which traverse the equilibrium quantum critical points from different directions and exhibit varying numbers of DQPTs. Figure~\ref{fig:S_r_t}(a1) shows the tr-RIXS spectra at $q=\pi/16$ during an $h$ pump that transitions the system from an ordered phase to a disordered phase and back. As shown in Fig.~\ref{fig:S_r_t}(a2), the system traverses one DQPT before returning to a state close to the pre-pump one, as indicated by the near-zero value of the post-pump $l(t)$. The constant energy cut at $\omega=0.38$ intersects the lower boundary of the two-kink continuum ($I_{++}$), $\omega_\mathrm{low}$, and the $\omega_\mathrm{diff}$ excitation ($I_{+-}$), where majority of the interference excitation $I_\mathrm{int}$ resides, and exhibits an intensity drop at the DQPT, consistent with the behavior observed in Fig.~\ref{fig:S_omega_t}. This correspondence becomes even more evident in the energy-integrated tr-RIXS, or tr-RXS $I_0(q, t_\mathrm{pr})$ [Eq.~\eqref{eq:trrixs:trrxs}], as shown in Fig.~\ref{fig:S_r_t}(a3).
Similarly, the Fourier-transformed tr-RXS into real space, shown in Fig.~\ref{fig:S_r_t}(a4), exhibits a clear intensity drop in the local equal-time correlator $\langle S^z (t_\mathrm{pr})^2 \rangle$ at the DQPT, reinforcing this connection.

Figures~\ref{fig:S_r_t}(b1)-(b4) present another example, starting in the disordered phase, where a $J$ pump drives the system into the ordered phase at its peak. Similar to Figs.~\ref{fig:S_r_t}(a1)-(a4), the system crosses one DQPT. However, it transitions to a final state markedly different from its initial state, as evidenced by the relatively large post-pump value of $l(t)$. The constant energy cut at $q = \pi/8$ and $\omega = 0.86$ in Fig.~\ref{fig:S_r_t}(b2) consistently shows an intensity drop at the DQPT, underscoring the robustness of this correspondence.
Notably, as the system parameters near the DQPT align with the equilibrium ordered state where the two-kink continuum exhibits prominent intensity [Fig.~\ref{fig:S_r_t}(b1)], resulting in the integrated intensity peaking around the DQPT. This behavior contrasts with Figs.~\ref{fig:S_r_t}(a1)-(a4), where the system transitions into a disordered phase during the pump. Consequently, the DQPT-related intensity modulation arising from the interference term $I_\mathrm{int}$ appears visually less pronounced in Fig.~\ref{fig:S_r_t}(b3) compared to Fig.~\ref{fig:S_r_t}(a3). A similar contrast is observed between Figs.~\ref{fig:S_r_t}(b4) and~\ref{fig:S_r_t}(a4), where a maximum of $\langle S^z(t_\mathrm{pr})^2 \rangle$ is observed at the DQPT. The validity of this correspondence extends to more complex scenarios where the system crosses the equilibrium quantum critical point multiple times, exhibiting multiple DQPTs. An illustrative example is shown in Figs.~\ref{fig:S_r_t}(c1)-(c4), using a few-cycle $J$ pump. The resulting spectra exhibit characteristics similar to those observed for the $h$ pump in Figs.\ref{fig:S_r_t}(a1)-(a4), with the correspondence between spectral features and DQPTs clearly discernible at each DQPT. These results collectively demonstrate that the correspondence between tr-RIXS spectral features and DQPTs is robust across diverse pump configurations and system trajectories, highlighting its potential as a broadly applicable diagnostic tool for DQPT.

\section{Conclusion}\label{sec:conclusion}

In this work, we investigated the nonequilibrium dynamics of TFIC with time-dependent magnetic exchange $J$ or external field $h$, using tr-RIXS as a probe in experimentally relevant pump-probe setups. Through a comprehensive analysis of equilibrium and nonequilibrium tr-RIXS spectra, we demonstrated their ability to capture the instantaneous parameters of the system’s Hamiltonian and to identify equilibrium quantum critical points, even during nonequilibrium dynamic evolution. Notably, we uncovered a striking one-to-one correspondence between DQPTs and oscillatory features in the low-energy tr-RIXS spectra, providing a practical and experimentally accessible approach to detect DQPTs. These findings highlight the versatility of tr-RIXS in probing quantum systems far from equilibrium, bridging the gap between theoretical predictions and experimental observations. Before closing, we briefly comment on the experimental relevance of our findings. Prominent candidate materials for realizing the TFIC include cobalt-based quasi-one-dimensional magnets~\cite{Yoshizawa1979, coldea2010, Bera2014, wang2016confined, Wang2018, amelin2020experimental}. The current energy resolution of RIXS at the $L$-edge of 3$d$ transition metals is around 10 meV~\cite{Yamamoto2025NanoTerasu}, and continues to improve with ongoing advances in instrumentation. This is already close to the exchange coupling $J$ in compounds such as CsCoCl$_3$\cite{Yoshizawa1979}, where $J \approx 7$ meV. The pump and probe timescales explored in this work, determined by $1/J$, correspond to approximately 100 femtoseconds, well within the capabilities of existing tr-RIXS setups~\cite{Merzoni2025}.
Taken together, our work opens new opportunities for studying nonequilibrium dynamics in condensed matter systems, with the exploration of spectral characterizations of DQPTs and other out-of-equilibrium phenomena in general quantum systems representing an intriguing direction for future research.

\begin{acknowledgments}
This work was supported by the National Key R\&D Program of China (No. 2022YFA1403000) and the National Natural Science Foundation of China (No. 12274207).
\end{acknowledgments}

\appendix
\section{Analysis of the tr-RIXS cross section}\label{app:crosssection}

To provide a comprehensive analysis of the tr-RIXS cross section~\eqref{eq:trrixs:crsec3}, we dissect its components in detail in this section. To simplify the analysis and enable a clearer interpretation of the tr-RIXS spectra, we approximate $J(t)$ or $h(t)$ as constant, taking values $J(t_\mathrm{pr})$ or $h(t_\mathrm{pr})$ within a small time window $[t_\mathrm{pr} - \delta t, t_\mathrm{pr} + \delta t]$ that spans the duration of the probe pulse. Within this time window, the time-dependent Bogoliubov evolve following Eq.~\eqref{eq:time bogo coeff}
\begin{eqnarray}
\mu_k(t_\mathrm{pr}+\dd t) &=& e^{i\omega_k \dd t }[v_k^2\mu_k(t_\mathrm{pr})+u_kv_k\nu_{-k}^*(t_\mathrm{pr})] \\
&& + e^{-i\omega_k \dd t}[u_k^2\mu_k(t_\mathrm{pr})-u_kv_k\nu_{-k}^*(t_\mathrm{pr})], \nonumber\\
\nu_{-k}^*(t_\mathrm{pr}+\dd t) &=& e^{i\omega_k \dd t}[u_k^2\nu_{-k}^*(t_\mathrm{pr})+u_kv_k\mu_k(t_\mathrm{pr})] \\
&& + e^{-i\omega_k\dd t}[v_k^2\nu_{-k}^*(t_\mathrm{pr})-u_kv_k\mu_k(t_\mathrm{pr})], \nonumber
\end{eqnarray}
after a small time step $\dd t$. Here, $\pm \omega_k $ are the instantaneous eigenvalues of $H(t)$, with corresponding instantaneous Bogoliubov coefficients $u_k$ and $v_k$. The tr-RIXS cross section \eqref{eq:trrixs:crsec3} can be decomposed into the following terms:
\begin{widetext}
\begin{align}
\label{eq:Ipp}
I_{++} = \frac{1}{4\pi\sigma_\mathrm{pr}^2N} \sum_k &e^{-\frac{1}{2}(\omega-\omega_k-\omega_{k+q})^2\sigma_\mathrm{pr}^2}g_{2,k}g_{2,k+q} (u_{k+q}^2v_k^2+u_{k+q}v_{k+q}u_kv_k),\\
\label{eq:Ipm}
I_{+-} = \frac{1}{4\pi\sigma_\mathrm{pr}^2N}\sum_k &e^{-\frac{1}{2}(\omega+\omega_k-\omega_{k+q})^2\sigma_\mathrm{pr}^2}g_{1,k}g_{2,k+q}(u_{k+q}^2u_k^2-u_{k+q}v_{k+q}u_kv_k) \\ \nonumber
+&e^{-\frac{1}{2}(\omega-\omega_k+\omega_{k+q})^2\sigma_\mathrm{pr}^2}g_{1,k+q}g_{2,k}(v_{k+q}^2v_k^2-u_{k+q}v_{k+q}u_kv_k),\\
\label{eq:Imm}
I_{--} =\frac{1}{4\pi\sigma_\mathrm{pr}^2N} \sum_k &e^{-\frac{1}{2}(\omega+\omega_k+\omega_{k+q})^2\sigma_\mathrm{pr}^2}g_{1,k}g_{1,k+q} (v_{k+q}^2u_k^2+u_{k+q}v_{k+q}u_kv_k),\\
\label{eq:Iint}
I_\mathrm{int} = \frac{1}{4\pi\sigma_\mathrm{pr}^2N}\sum_k & [e^{-\frac{1}{2}[(\omega-\omega_{k})^2+\omega_{k+q}^2]\sigma_\mathrm{pr}^2}g_{2,k}-e^{-\frac{1}{2}[(\omega+\omega_{k})^2+\omega_{k+q}^2]\sigma_\mathrm{pr}^2}g_{1,k}] \\ \nonumber
\times & 2 \Re(g_{3,k+q}) [(u_{k+q}^2-v_{k+q}^2)u_kv_k+u_{k+q}v_{k+q}(u_k^2-v_k^2)], \\
\label{eq:Iint2}
I_\mathrm{int2} = \frac{1}{4\pi\sigma_\mathrm{pr}^2N}\sum_k & e^{-\frac{1}{2}[(-\omega_k+\omega_{k+q})^2+\omega^2]\sigma_\mathrm{pr}^2}g_{3,k+q}g_{3,k}^* (u_{k+q}u_k-v_{k+q}v_k)^2 \\ \nonumber
 - & e^{-\frac{1}{2}[(\omega_k+\omega_{k+q})^2+\omega^2]\sigma_\mathrm{pr}^2}\times  2\Re(g_{3,k+q}g_{3,k})(u_{k+q}^2v_k^2+u_{k+q}v_{k+q}u_kv_k).  \\ \nonumber
\end{align}
\end{widetext}
The coefficients are defined as
\begin{equation}
\begin{split} \label{eq:g1k}
g_{1,k} &= |  \langle \psi_k(t_\mathrm{pr}) | \psi_k^+   \rangle   |^2 \\
&=v_k^2\mu_k(t_\mathrm{pr})\mu_k^*(t_\mathrm{pr})+u_k^2\nu_k(t_\mathrm{pr})\nu_k^*(t_\mathrm{pr})  \\
&\quad -u_kv_k[\nu_k^*(t_\mathrm{pr})\mu_k^*(t_\mathrm{pr})+\mu_k(t_\mathrm{pr})\nu_k(t_\mathrm{pr})],
\end{split}
\end{equation}
\begin{equation}
\begin{split} \label{eq:g2k}
g_{2,k} &= |  \langle \psi_k(t_\mathrm{pr}) | \psi_k^-   \rangle   |^2 \\
&=u_k^2\mu_k(t_\mathrm{pr})\mu_k^*(t_\mathrm{pr})+v_k^2\nu_k(t_\mathrm{pr})\nu_k^*(t_\mathrm{pr})  \\
&\quad +u_kv_k[\nu_k^*(t_\mathrm{pr})\mu_k^*(t_\mathrm{pr})+\mu_k(t_\mathrm{pr})\nu_k(t_\mathrm{pr})],
\end{split}
\end{equation}
\begin{equation}
\begin{split} \label{eq:g3k}
g_{3,k}&=-v_k^2\nu_k^*(t_\mathrm{pr})\mu_k^*(t_\mathrm{pr})+u_k^2\mu_k(t_\mathrm{pr})\nu_k(t_\mathrm{pr})  \\
&\quad +u_kv_k[\nu_k(t_\mathrm{pr})\nu_k^*(t_\mathrm{pr})-\mu_k(t_\mathrm{pr})\mu_k^*(t_\mathrm{pr})],
\end{split}
\end{equation}
Here, $g_{1,k}$ and $g_{2,k}$ quantify the weight of the time-evolved wave function in the instantaneous excited ($|\psi^+\rangle$) and ground ($|\psi^-\rangle$) states, respectively. They satisfy the normalization condition $g_{1,k}+g_{2,k}=1$. Furthermore, it can be shown that $(g_{1,k}-\frac{1}{2})^2+ | g_{3,k} | ^2=\frac{1}{4}$.

\begin{figure}[tb]
    \centering
    \includegraphics[width=\linewidth]{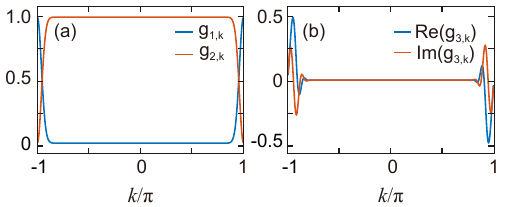}
    \caption{\label{fig:gk} (a) Momentum dependence of $g_{1,k}$ (blue) and $g_{2,k}$ (red). (b) Momentum dependence of $\Re (g_{3,k})$ (blue) and $\Im (g_{3,k})$ (red). }
\end{figure}

\begin{figure}[tb]
    \centering
    \includegraphics[width=\linewidth]{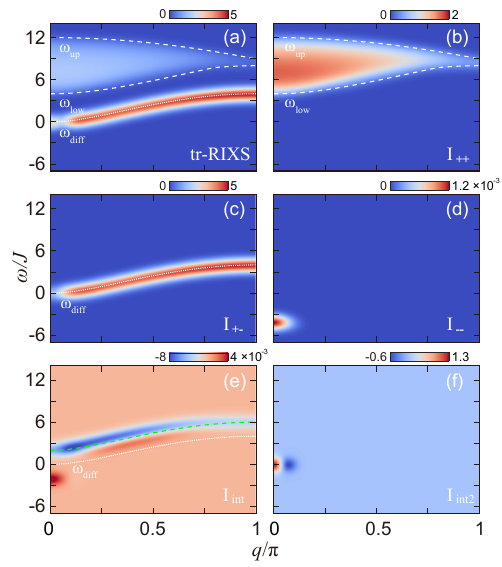}
    \caption{\label{fig:split} Decomposition of the nonequilibrium tr-RIXS spectra. (a) The tr-RIXS spectra corresponding to Fig.~\ref{fig:noneq}(c) under the approximation $h(t) \approx h(t_\mathrm{pr})$ during the probe. (b)-(f) The five decomposed components of the full tr-RIXS cross section: $I_{++}$, $I_{+-}$, $I_{--}$, $I_\mathrm{int}$, and $I_\mathrm{int2}$, respectively. The dashed lines in (a) and (b) indicate the boundaries of the instantaneous two-kink continuum determined by $H(t_\mathrm{pr})$. The white dotted line in (a), (c), and (e) represents $\omega_\mathrm{diff}$. The green dashed line in (e) denotes  $\omega_{\pi-q}$.}
\end{figure}

Figure~\ref{fig:gk} shows the momentum dependence of these coefficients for the pump configuration in Fig.~\ref{fig:noneq}(c). While the exact line shapes vary with different pump configurations, some general qualitative trends can be expected. Since single-particle excitations are most easily induced around $k = \pi$, where the gap is minimal, $g_{1,k}$, which characterizes the weight of the excited mode $| \psi_k^+   \rangle $, remains small except near $k=\pi$. This simultaneously determines the behavior of $g_{2,k}$, which remains close to 1 except near $k=\pi$. Likewise, $g_{3,k}$, which shares the same time-dependent Bogoliubov coefficients with $g_{1,k}$ and $g_{2,k}$, is expected to exhibit significant variations primarily around $k=\pi$.

Figure~\ref{fig:split} shows the various components of the cross section. The component $I_{++}$ generates the two-kink continuum, analogous to the equilibrium case, with intensities primarily distributed within range $\omega = \omega_k + \omega_{k+q}$, as dictated by the exponential factor in Eq.~\eqref{eq:Ipp}.
$I_{+-}$ gives rise to the $\omega_\mathrm{diff}$ excitation as discussed in the main text. Instead of a continuum, its intensity is localized along a single dispersion line associated with quasiparticle excitations around $k=\pi$, where the energy gap is smallest. This is also clear from the coefficient $g_{1,k}$, which reach its maximum at $k=\pi$ in Fig.~\ref{fig:gk}(a).
The term $I_{--}$ is expected to give rise to a continuum analogous to $I_{++}$ but with opposite energy distribution $\omega = -\omega_k - \omega_{k+q}$. However, its intensity is predominantly localized near  $q = 0$. This behavior arises for reasons analogous to those governing  $I_{+-}$. Since $g_{1,k}$ is effectively zero except near $k = \pi$, the product $g_{1,k}g_{1,k+q}$ in Eq.~\eqref{eq:Imm} ensures that  $I_{--}$ has nonzero intensity only near $q = 0$, involving two quasiparticles with momenta close to $k = \pi$ and a combined energy of approximately $-4$.
There are two additional terms, $I_\mathrm{int}$ and  $I_\mathrm{int2}$, which do not lend themselves to a straightforward interpretation as two-quasiparticle excitations. These terms likely capture more complex contributions arising from interference of different excitation channels, reflecting the nontrivial nature of the nonequilibrium dynamics. However, their intensity distribution can still be inferred from Eqs.~\eqref{eq:Iint} and \eqref{eq:Iint2}. In Eq.~\eqref{eq:Iint}, the term $e^{-\frac{1}{2}[(\omega-\omega_k)^2+\omega_{k+q}^2]\sigma_\mathrm{pr}^2}g_{2,k}$, where the exponential factor peaks at $\omega = \omega_k$, contributes to a widespread intensity across momentum space, as $g_{2,k}$ remains close to 1 over most of the momentum range, resulting in a broadly distributed negative contribution. This term is further multiplied with $g_{3,k+q}$, which exhibits sharp maxima near $k = \pi$ [Fig.~\ref{fig:gk}(b)]. This results in a negative intensity distribution along $\omega = \omega_{\pi-q}$. By a similar argument, the intensity related to the term $e^{-\frac{1}{2}[(\omega+\omega_k)^2+\omega_{k+q}^2]\sigma_\mathrm{pr}^2}g_{1,k}$ is localized near the energy-momentum point $\omega = -\omega_\pi=-2$, $q=0$. For the component $I_\mathrm{int2}$ defined in Eq.~\eqref{eq:Iint2}, the first exponential factor is maximized near $\omega=0$ and $q = 0$, while the second exponential factor is generally much smaller and contributes negligibly when a quasiparticle gap exists. At $q=0$, the weight factor $g_{3,k+q}g_{3,k}^*$ is maximized for two $k$ points near $\pi$, as shown in Fig.~\ref{fig:gk}(b). For a small but nonzero $q$, the two $k$ points with maximal $g_{3,k}$ amplitude but opposite signs can be connected, resulting in a dominant negative contribution from $g_{3,k+q}g_{3,k}^*$. This interplay produces the two prominent intensity distributions in Fig.~\ref{fig:split}(f): one at $q = 0$ and another at a small, positive $q$, exhibiting opposite intensities.

\section{Correspondence between low-energy spectral features and DQPT}\label{app:correspondence}

\begin{figure}[tb]
    \centering
    \includegraphics[width=\linewidth]{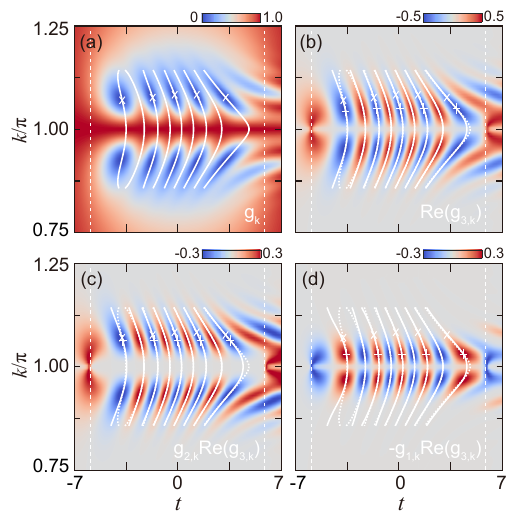}
    \caption{\label{fig:correspondence} The momentum and time distribution of (a) $g_{k}$, (b) $\Re(g_{3,k})$, (c) $g_{2,k}\Re(g_{3,k})$, and (d) $-g_{1,k}\Re(g_{3,k})$ during a $h$ pump corresponding to Figs.~\ref{fig:noneq}(a)-(e). Although $\Re(g_{3,k})$ is odd in $k$, it is plotted as an even function because the finial result $I_\mathrm{int}$ is even. In each panel, the dashed lines indicate the time when $h(t)/J = 1$, and the solid lines show the extrema of each plotted quantity. In panels (b)-(d), the extrema of $g_{k}$ are also shown as dotted lines for comparison. The cross marks ($\times$) indicate the positions where $g_{k}=0$, while the plus signs ($+$) denote the minima of $\Re (g_{3,k})$, $g_{2,k}\Re (g_{3,k})$, or the maxima of $-g_{1,k}\Re (g_{3,k})$.}
\end{figure}

The low-energy spectral weight oscillations primarily arise from the interference term $I_\mathrm{int}$ [Eq.~\eqref{eq:Iint}] in App.~\ref{app:crosssection}. The instantaneous Bogoliubov coefficients $u_k$ and $v_k$ exhibit a smooth and gradual variation over time, except when crossing the equilibrium quantum critical point. Thus, the oscillatory behavior is potentially driven by the coefficients $g_{1,k}$, $g_{2,k}$, and $\Re(g_{3,k})$ [Eqs.~\eqref{eq:g1k}-\eqref{eq:g3k}], which characterize the time-dependent wave function $|\psi(t)\rangle$ in relation to the system’s instantaneous eigenstates. Taking their time derivatives reveals that the intensity extrema are primarily determined by $\Re(g_{3,k})$. Neglecting the time derivatives of the instantaneous Bogoliubov coefficients $u_k$ and $v_k$, the derivative of $\Re(g_{3,k})$ is approximately given by a relatively simple expression
\begin{equation}
\begin{split} \label{eq:dg3k}
\frac{\mathrm{d} \Re(g_{3,k})}{\mathrm{d} t} &\approx[\alpha_k(t)\delta_k(t)+\beta_k(t)\gamma_k(t)] \\
&\quad \times 2\omega_k [(u_k^2-v_k^2)^2+4u_k^2v_k^2],
\end{split}
\end{equation}
where $\mu_k(t)$ and $\nu_k(t)$ are expressed as their real and imaginary parts, $\mu_k(t)=\alpha_k(t)+ \ii \beta_k(t)$ and $\nu_k(t)=\gamma_k(t)+\ii \delta_k(t)$. To explore the correspondence between the intensity extrema and the DQPT, we further investigate the $k$-resolved return probability $g_k= | \langle \psi_k^i | \psi_k(t) \rangle |^2$ [see Eq.~\eqref{eq:ltTFIC}]. A nonanalyticity in $l(t)$ corresponds to a zero in $g_k(t)$ at certain $k$. Since $g_k(t)$ is nonnegative, these zeros occur at the minima, which can be located by finding the zeros of its time derivative
\begin{equation}
\begin{split} \label{eq:dgk}
\frac{\mathrm{d} g_{k}}{\mathrm{d} t} &=[\alpha_k(t)\delta_k(t)+\beta_k(t)\gamma_k(t)] \\
&\quad \times 4\omega_k \{(u_k^2-v_k^2)u_k^iv_k^i-[(u_k^i)^2-(v_k^i)^2]u_kv_k \},
\end{split}
\end{equation}
where $u_k^i$, $v_k^i$ are the initial Bogoliubov coefficients. Note that in Eq.~\eqref{eq:dgk}, the expression inside the curly brackets vanishes only when the Hamiltonian returns to its initial parameters (see Sec.~\ref{sec:method:corfun}). Equations~\eqref{eq:dg3k} and~\eqref{eq:dgk} share a common time-dependent factor, and the extrama of $\Re(g_{3,k})$ and $g_k$ are both found by locating the derivative zeros determined by $\alpha_k(t)\delta_k(t) + \beta_k(t)\gamma_k(t) = 0$. A nuance exists: while the potential zeros at minima of $g_k(t)$ [Eq.~\eqref{eq:dgk}] at a specific momentum $k$ directly signal a DQPT, the intensity extrema do not directly result from the zeros of Eq.~\eqref{eq:dg3k}, as they involves an integration over the entire momentum space [Eq.~\eqref{eq:Iint}], which may break the correspondence. Based on the analysis in App.~\ref{app:crosssection}, $g_{3,k}$ only acquires finite values close to $k=\pi$, a momentum corresponding to the gap minimum of the single-particle excitations. Therefore, the integration does not necessarily break the correspondence, as long as the zeros in Eq.~\eqref{eq:dg3k} do not exhibit strong $k$-dependence around $k=\pi$.

Figure~\ref{fig:correspondence} shows the momentum and time distribution of $g_k$ and $\Re(g_{3,k})$ around $k=\pi$, using pump configurations identical to those in Figs.~\ref{fig:noneq}(a)-(e). Within the time window $-5.9<t<5.9$, where the DQPTs occur, the extrema of $g_k$ and $\Re(g_{3,k})$ generally exhibit relatively weak $k$ dependence around $k=\pi$ and show strong agreement in Figs.~\ref{fig:correspondence}(a) and~\ref{fig:correspondence}(b). A notable $k$ dependence is observed only for the last extremum, which is likely due to its close proximity to the equilibrium quantum critical point and also the rapid change of $h(t)$ at that time. The additional factors $g_{1,k}$ and $g_{2,k}$ further modulate the intensities corresponding to the interference term $I_\mathrm{int}$ [Eq.~\eqref{eq:Iint}] at positive and negative energies, respectively. Since their variations within the time window are relatively smooth, the resulting extrema distribution shows only minimal modification compared to $\Re(g_{3,k})$ alone, as shown in Figs.~\ref{fig:correspondence}(c) and \ref{fig:correspondence}(d). A small difference between the effects of $g_{1,k}$ and $g_{2,k}$ is that the former shifts the minima of the positive-energy intensity closer to the DQPTs in time, while the latter shifts the maxima of the negative-energy intensity in the opposite direction. This accounts for the relatively better correspondence at positive energies.

Finally, we note that since the correspondence derived here does not explicitly involve the energy $\omega$, it is naturally expected to hold for the energy-integrated tr-RIXS cross section, or tr-RXS. In addition, we emphasize that although the above derivations are based on the exactly solvable TFIC, numerical simulations indicate that the correspondence persists in the presence of weak integrability-breaking perturbations such as small next-nearest-neighbor interactions.

%

\end{document}